\documentclass[aps,pre,amssymb,amsmath,twocolumn,floatfix]{revtex4-2}
\usepackage{etoolbox}
\usepackage{breqn}
\usepackage{graphicx}
\usepackage{dcolumn}
\usepackage{bm}
\usepackage{subcaption}

\makeatletter
\preto\maketitle{%
	\begingroup\lccode`~=`,
	\lowercase{\endgroup
		\let\saved@breqn@active@comma~
		\let~}\active@comma 
}
\appto\maketitle{%
	\begingroup\lccode`~=`,
	\lowercase{\endgroup
		\let~}\saved@breqn@active@comma 
}
\makeatother

\graphicspath{ {./img/} }
\begin{document}

\title{XY model on a self-avoiding walk }

\author{Kamilla Faizullina}
\author{Evgeni Burovski}
\affiliation{HSE University, 101000 Moscow, Russia}

\begin{abstract}
	We study a lattice model of a magnetic polymer where the XY spin variables are located on a self-avoiding walk (SAW) on a regular lattice in two and three dimensions.
	We consider the regime where both spins and conformations are dynamic, thus the XY model is defined on a dynamic lattice and conformations generate an annealed disorder. Using Monte Carlo simulations, we characterize the globule-coil and ferromagnetic phase transitions, and pay special attention to the vicinity of the theta-point. Our numerical results suggest that the transitions are continuous in two dimensions and first-order in three dimensions, which is similar to related models with Ising spins.  
\end{abstract}

\maketitle

\section{\label{sec:level1}Introduction}

Coarse-grained modelling is a commonly used approach to study structural properties of polymeric materials \cite{deGennes1979}.   
One of the  models of a macromolecule is a linear polymer which is represented by an interacting (also known as collapsing) self-avoiding walk (SAW). Self-avoiding walks allows to include excluded volume effects for polymers in a good solvent. Van der Waals type attraction is modelled via including the nearest neighbour monomer attraction. Critical phenomena take place in the infinite systems in second order phase transition which is defined as a singularity of free-energy function (see Chap.3 in Ref \cite{phasetransition}).

In the simplest case, a homopolymer chain consists of one type monomers. The polymers with different types of subunits are called heteropolymers. The simplest heteropolymer model is Hydrophobic-polar (HP) model of protein \cite{lau1989lattice}. It assumes that the sequence of monomers types are fixed. This model was introduced to approximate the folding process of protein and mostly used for development algorithms to find minimum energy states (for example, \cite{fress, PhysRevE.68.021113}). This model also was used to explore conformations space of proteins \cite{HELLING2001157}. Recently,  the dynamical case of HP model was studied \cite{Faizullina2021}. In that dynamics, sequence of monomers and geometry structure are not fixed . Computational results are consistent with the assumption that dynamical HP model  and an interacting homopolymer have the similar behaviour in phase transition point and they are in the same universality class at the conformational transition point.

To represent the ferromagnetic properties of polymer materials self-avoiding walk models with spin dynamics were introduced \cite{Garel1999}. Ising model is the simplest example of system which undergoes a phase transition between ordered and disordered states. These models  were studied it for 2D and 3D lattices \cite{Garel1999, Papale2018} using mean-field theory and Monte-Carlo simulations. 
This model was studied  using Monte-Carlo methods for long chains  \cite{PhysRevE.104.024122, PhysRevE.104.054501}  on regular square lattices. 
Computational results show that  the transition between swollen and collapsed phases is second-order in 2D case and first-order magnetic transition on 3D lattice with undergoing swollen/collapsed states transition. 

Recently, magnetic polymers were studied on Sierpiński triangle in 3D \cite{PhysRevE.108.L042502}, where renormalisation group approach was used to present phase diagram showing magnetic and conformational phases. 

The other example of magnetic polymer is Potts-like model on SAWs. This model in case of regular lattice was  introduced to represent vulcanization \cite{AConiglio_1979}. Recently, Potts-like model was studied for  Bethe lattice  \cite{PhysRevE.106.024130}.  

Previous work was for Ising spins, and in this work, we extend to XY model where spin variables are continuous.  Original XY model on regular 2D square lattice has a topological order, which was proposed theoretically and named Kosterlitz-Thouless (KT) phase transition \cite{Kosterlitz_1973}. As other spin models, classical 2D was studied numerically using Monte-Carlo methods \cite{xy2005, nikolaou2007matter}. 

In this work, we explore XY model on SAWs for 2D and 3D lattices in lack of an external field. We construct Monte-Carlo algorithm to study this system at the phase transition region.

 
\section{Model and method}
\textit{Model.} 
A polymer conformation of the length $N$ is a self-avoiding walk (SAW) with $N-1$ edges and $N$ nodes on a regular lattice.  
Each $i$th node represents a spin-like variable $s_i$ which is associated with angle $\theta_i \in  [-\pi;\pi)$. 

The Hamiltonian for  a sequence of spins, $s$, and a conformation, $u$, is defined as the sum over all  non-repeating neighbour  pairs $\langle i, j \rangle$th in conformation:
\begin{dmath}
\label{hamiltonian}
H(u,s) = -J \sum_{ \langle i, j \rangle } \cos(\theta_i - \theta_j) - h \sum_i \cos(\theta_i)
\end{dmath}
Here, $J>0$ is the coupling constant  which represents spin-spin attraction. 
In our work, we focus on the system in lack of an external field: $h=0$.  
Without loss of generality, we assume that $\beta = \frac{1}{kT} = 1$, where $k$ is Boltzmann’s constant, $T$ is temperature. 

Let $U_N$ be a set of all SAW conformations of N monomers.
The partition function for the chain of the length $N$ is the sum over all SAW conformations $u \in U_N$  of N monomers and the integral over all spin space:  
\begin{multline}
\label{partitionfunction}
Z(J) =  \sum_{u \in U_N }  \int_{-\pi}^{\pi}   \frac{1}{ (2 \pi  )^N}
e ^{J \cos(\theta_1-\theta_2)} e ^{J \cos(\theta_2-\theta_3)} \dots  \\ 
e ^{J \cos(\theta_{N-1}-\theta_N)}    d \theta_1 d \theta_2 \dots d\theta_N  \;.
\end{multline}

\textit{Physical observables}.
The magnetization is defined as a vector:
\begin{dmath}
\label{meanmagnetization}
\langle \vec{m} \rangle =    \frac{1}{N} \left\langle ( \sum_{i=1}^{N} \cos \theta_i, \sum_{i=1}^{N} \sin \theta_i  ) \right\rangle \;,
\end{dmath}
where $\langle... \rangle$ is averaging  with respect to the Gibbs distribution  \eqref{partitionfunction}. 
The second moment of magnetization is a square of the norm:
\begin{dmath}
\label{secondmomentmagnetization}
\langle m^2 \rangle = \frac{1}{N^2} \left\langle ( \sum_{i=1}^{N} \cos \theta_i )^2 +  (\sum_{i=1}^{N} \sin \theta_i  )^2 \right\rangle \;.
\end{dmath}

From measurements of the average magnetization per spin $\langle m \rangle (J)$, we can obtain the value of the magnetic cumulant  (Binder parameters) of fourth order \cite{Binder2010}, which is helpful to study magnetic phase transition:
\begin{dmath}
\label{binderqum}
U_4 (J) = 1 - \frac{ \langle m^4 \rangle}{3 \langle m^2 \rangle^2  } \;.
\end{dmath}
At a
continuous phase transition, the Binder ratio transitions approaches a step function as the system size is
increased. The Binder ratio has a divergent feature at the step if the system has the first order transition. At the ordered phase $J>J_{cr}$, the value $U_4 \rightarrow 2/3$. At the disordered state $J \rightarrow 0$,  $U_4 \rightarrow 1/3$  (see in Sec. \ref{u4xy}).  

To study structural phase transition, we use the mean square end-to-end distance (radius) of self-avoiding-walks which is defined as the sum over all configurations:
\begin{dmath}
\label{endtoend}
\langle R_N^2 \rangle  =  \frac{1}{Z_N} \sum_{ u \in U_N }  |u|^2 e^{-H(s,u)} \;,
\end{dmath}
where $|u|$ is the Euclidean distance between the endpoints of conformation $u$, and $Z_N$ is partition function \eqref{partitionfunction}. We call it "mean radius" for brevity. As $N \rightarrow \infty $, the mean radius  of SAWs is believed to scale as 
\begin{dmath}
\label{r_scale}
\langle R_N^2 \rangle \sim N^{2 \nu }.
\end{dmath}
Here ${\nu} $ is a critical exponent. For thermodynamic limit $N \rightarrow \infty $, $\nu$ is believed to have the form of a step function of interaction energy $J$. For finite systems, this effect is rounded \cite{vanderzande1998lattice}. This exponent ${\nu} $    defines three regimes:  swollen, theta and compact. 

 The exact value of critical  exponent for non-interacting SAWs ($J=0$) on the square lattice  \cite{Li1995}
\begin{dmath}
\label{nur}
\nu = \frac{3}{4}. 
\end{dmath}
At the theta-point, $\nu_{\theta}$  is obtained via Coulomb-gas method  \cite{Duplantier1987}:
\begin{equation}
\label{nu_theta}
\nu_{\theta} = \frac{4}{7} \approx 0.57 .
\end{equation}
For the globular regime ($J > J_{\theta}$) in 2D case:
\begin{equation}
\label{globular}
\nu = \frac{1}{d} = \frac{1}{2}.
\end{equation}

At low $J < J_{\theta}$, the system is equivalent to SAW without interaction. One should call to mind the classical homopolymer model which is represented by an interacting, or collapsing, self-avoiding walk (iSAW). Below we briefly report known critical values for homopolymer which are important in our work as iSAW is a parental model of XY on SAWs.

For 3D case, Flory predicted value for non-interacting self-avoiding walk as follows \cite{flory1953principles}:
\begin{equation}
\label{eq:nu_3D_J0_flory}
\nu = \frac{3}{2+d} = \frac{3}{5}.
\end{equation}

At the theta-point, the $\nu_{\theta}$  is \cite{van2015statistical} 
\begin{dmath}
\label{eq:nu_3D_Jtheta_flory}
\nu_{\theta} = \frac{1}{2}.
\end{dmath}

For compact regime when $J > J_{\theta}$, the critical exponent $\nu$  has following value:  
\begin{equation}
\label{eq:nu_3D_Jglobular_flory}
\nu = \frac{1}{d} = \frac{1}{3}.
\end{equation}
\textit{Method.}

In this work, we construct the Markov Chain Monte Carlo method for fixed-length chain consisting of three types of updates similar to \cite{PhysRevE.104.054501}.
We refer to them as BEE-reptation step, Reconnection and Wolff Cluster update. 
In each iteration, the algorithm chooses the update according to set of probabilities. 
We define these probabilities as $P_{\mathrm{local}}$,$P_{\mathrm{reconnect}}$ and $P_{\mathrm{Wolff}}$ respectively. 
The sum of probabilities is always equal to one: $P_{\mathrm{local}} + P_{\mathrm{reconnect}} + P_{\mathrm{Wolff}} = 1$. 
 
BEE-reptation move is a bilocal reptation update \cite{Caracciolo2002}. 
The algorithm removes a monomer from one end and adds a monomer to the other end.
Spin angle value $\theta_{new}$ and the direction in conformation are randomly generated.
The direction is chosen uniformly with the probability $1/(2d)$, where $d$ is the dimensions of the lattice.
The spin angle variable $\theta_{new}$ is generated uniformly $\theta_{new} \sim U(-\pi, \pi)$. 
The new generated state is simply accepted  according to the Metropolis rule:
 \begin{dmath}
 \label{accratios}
 A(u_0  \rightarrow u_{\text{new}} ) =  
 \begin{cases}
 e^{-J(E_{u_{\text{new}}}-E_{u_0})}, & \text{if $E_{u_{\text{new}}}-E_{u_0}>0$;}\\
 1, & \text{otherwise}.
 \end{cases}
 \end{dmath}
 
The BEE-reptation update has the time and the  memory complexity $O(1)$, however, the autocorrelation time is quite long for magnetic variables and for structure  $ \eta \sim N^2 $. The system can also be locked in the frozen states when both ends of the conformation are surrounded by $2d$ neighbours. 

To overcome disadvantages of bilocal move, we also use Reconnect update to accelerate conformations generation and Wolff-cluster algorithm to effectively explore spin configuration space. Reconnect is a non-local update based on ideas of Worm algorithm for Ising model \cite{Worm}.   In this system update, only connections of conformation are changed. The acceptance probability is always equal to one as the energy does not change. The time complexity $O(N)$.

The Wolff-cluster algorithm is a classical Monte Carlo simulation which allows to improve sampling of the spin configuration space  \cite{wolff}. 
The main idea of the update is to form the cluster of spins and flip its spins.
We follow the classical way of cluster update implementation for XY model discussed in Ref. \cite{newman1999monte}. 
The algorithm effectively sample spin configurations and keep the conformation fixed. The time complexity $O(N)$.

\section{Numerical simulations, 2D case} 

To study system on square lattice, we simulate chains up to $N=4900$. For $N=4900$, we run at least $8 \times 10^{10} $ MC steps. Here we use these values for update probabilities: $P_{\mathrm{local}}=0.8$, $P_{\mathrm{reconnect}}=0.199$, $P_{\mathrm{Wolff}}=0.001$. Despite both Reconnect and Cluster  updates  have complexity $O(N)$, we choose small $P_{\mathrm{Wolff}}$ due to slow iterations cause by using queue for creating cluster of spins. At average, the Reconnect update is much faster and leads to faster convergence of geometry properties, for example, mean radius. 

  \subsection{Structural properties}

    \begin{figure}
  	\centering 
  	\includegraphics[scale=0.25]{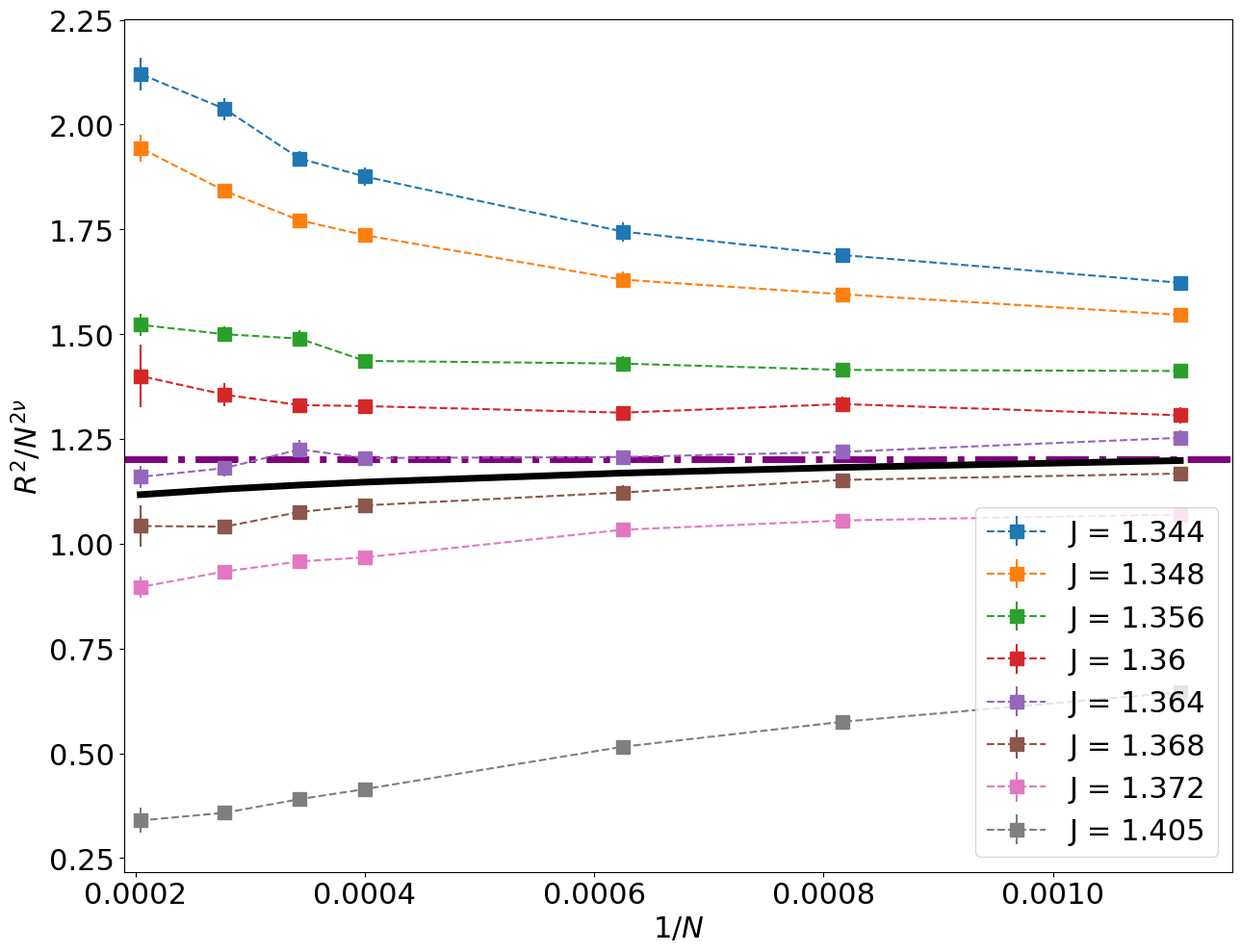}
  	\caption{The points are Monte-Carlo data for Mean-squared end-to-end distance scaled using $\nu_{\theta}=4/7$ \eqref{nu_theta}. The scaled radius is presented  as a function of $1/N$ from $N=900$ to $N=4900$.  Dashdotted purple line corresponds to  the structural transition with critical exponent $\nu_{\theta}=4/7$.  The  slope of black solid line corresponds to  $\nu=1/2$ from unfolded phase.  The dashed lines are guides for an eye. }
  	\label{fig:Rscaled_narrow}
  \end{figure}

  We start our studying structural properties of the model with analysing mean radius. From studies of dynamic HP model and Ising model on the SAWs which inherits the critical value $\nu_{\theta}=4/7$ \eqref{nu_theta} \cite{Faizullina2021,PhysRevE.104.054501,PhysRevE.104.024122}, we expect that XY model also inherits this value from parental model of interacting self-avoiding walk. 
  
  We make visual inspection of the scaling function for mean radius \eqref{endtoend}.  Here and elsewhere, we use estimations for $J_{\theta}$ that are obtained using paired regressions described in \ref{pairedregressions}.

  Figure \ref{fig:Rscaled_narrow} shows the scaled
  mean-squared end-to-end distance by $\nu=4/7$ as a function of 
  $1/N$ for different $J$. 
  The purple dash-dotted horizontal line is placed correspondingly to estimation $J_{\theta}$ from Section \ref{sec:2DJthetatransition}.  
  We also plot the value $\nu=1/2$ from high-temperature regime using black solid line with low slope. 
  We thus use assumption that XY model on SAWs also has value $\nu = 4/7$  \eqref{nu_theta} at the point of structural phase transition. 
  We use this value to obtain collapsing plots in Figure \ref{fig:bcshort} in following subsection \ref{section:Transition}.

  \subsection{Transition} \label{section:Transition}
\begin{figure}
  	\centering
  	\captionsetup{justification=centering}
  	\begin{subfigure}[b]{0.45\textwidth}
  		\includegraphics[width=\textwidth]{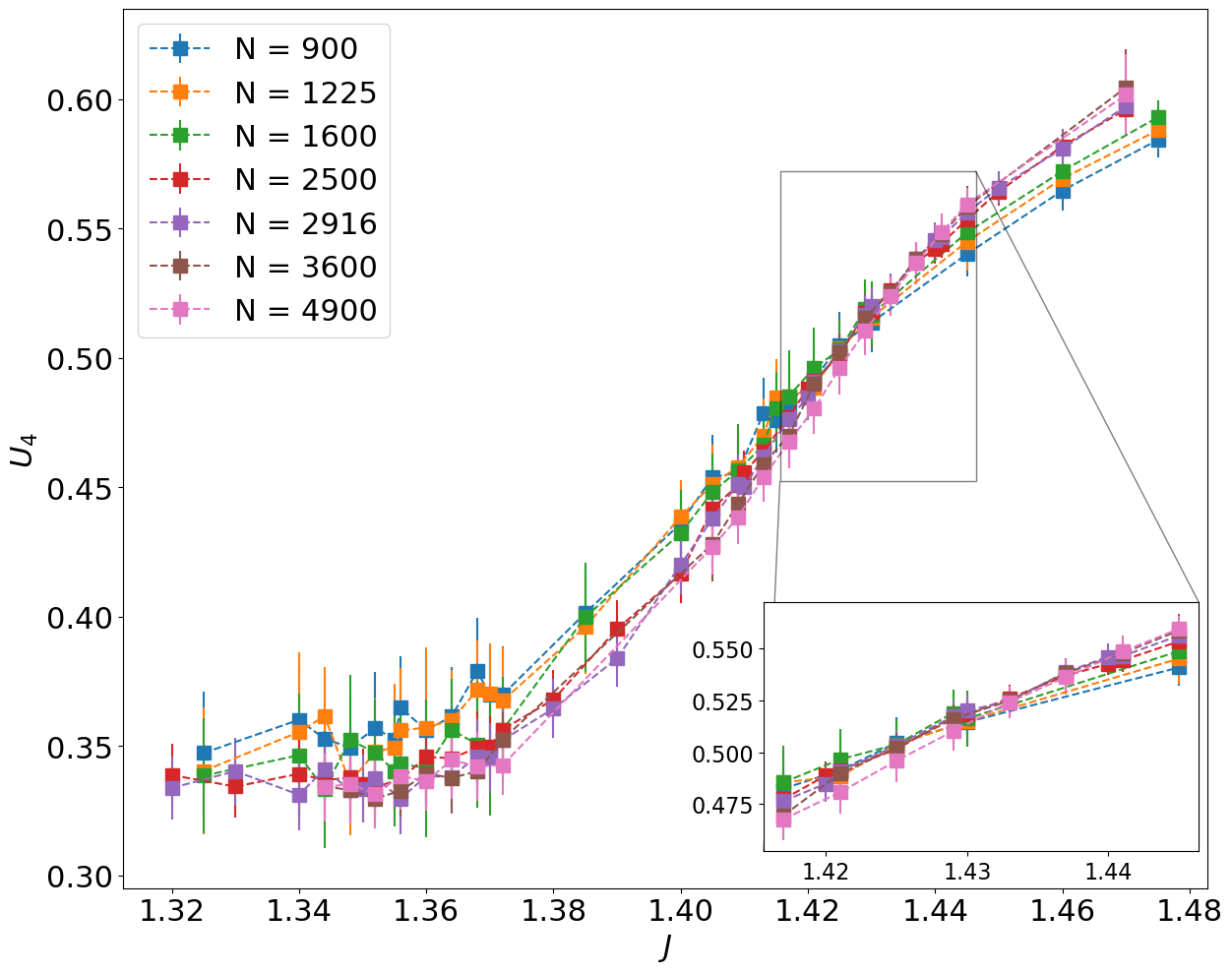}
  		\caption{ Binder cumulants for long chains across the critical region.}
  		\label{fig:bcshort_longbc}
  	\end{subfigure}
  	\begin{subfigure}[b]{0.45\textwidth}
  		\includegraphics[width=\textwidth]{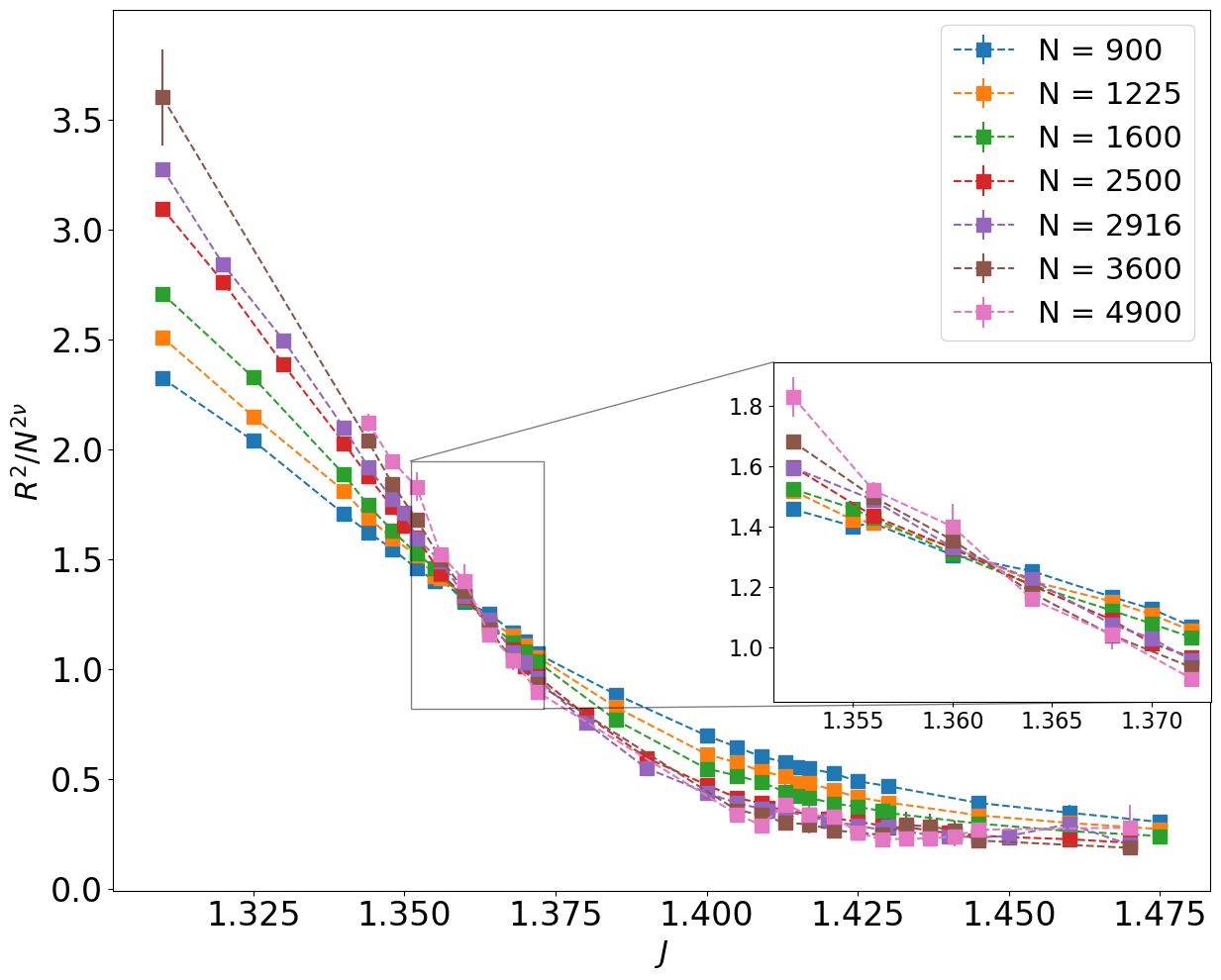}
  		\caption{Scaled mean radius with zoom-in at the critical interval }
  		\label{fig:bcshort_longradius}
  	\end{subfigure}
  	\caption{ Binder  cumulants \eqref{binderqum} and mean radius \eqref{endtoend} scaled by $\nu=\frac{4}{7}$.   }
  	\label{fig:bcshort}
  \end{figure}

  To focus on studying phase transition, we calculate two characteristics. The first one is the mean square end-to-end distance scaled using the factor $\nu =  4/7$ in \eqref{r_scale}. The second one is Binder cumulant of magnetization \eqref{binderqum}. Figure \ref{fig:bcshort}  presents obtained calculations. 
  
  \subsubsection{Magnetic phase transition} \label{sec:crossing}
  
  We compute Binder cumulants \eqref{binderqum} with the aim of determining  order of magnetic transition. 
  
  The Binder cumulants (top Figure \ref{fig:bcshort_longbc}) has lower limiting value $U_4=1/3$ as $J\rightarrow 0$ \eqref{binderqum_0} (see Sec. \ref{u4xy} for more details). For large interaction energy  constant $J$, the system  signs ordering behaviour $U_4=\frac{2}{3}$.

  Figure \ref{fig:bcshort_longbc} shows that Binder parameter curves does not diverge. Therefore, our computational results do not detect the first-order transition.

  To estimate critical values for cumulants $U_4$ and phase transition point $J_{cr}$, we perform paired linear intersections in the narrow regions where the curves cross (see \ref{pairedregressions}). We repeat  intersections for $N=3600$ with other chains. 
  Estimation of critical point is the   constant value from linear fit of the line obtained on pairs ($1/N_i, J_{cross}(N_i, 3600)$). The errorbars are calculated using module $linregress$ from SciPy \cite{2020SciPy-NMeth}. We obtain the following results: 
  \begin{dmath}
  \label{eq:critical_J_magnet_2D}
  J_{cr}^{3600} \approx  1.435 \pm 0.008
  \end{dmath}

  The estimation of critical value of cumulant at the magnetic transition from paired regressions  is following: $U_{4\mathrm{critical}}\approx 0.55(5) $. This value is far from the Binder cumulant  value for classical Ising model on the square lattice with periodic boundary condition and for Ising model on SAWs in 2D \cite{PhysRevE.104.054501}.

  \subsubsection{ Estimation of $\hat{J_{\theta}}$} \label{sec:2DJthetatransition}
  
 Figure \ref{fig:bcshort} shows that curves of mean radius for range of $N$ values cross approximately at the same point.
   
  Applying paired regressions method, we obtained following estimated value from the zero point:
  \begin{dmath}
  \label{eq:critical_J_theta_2D}
  J_{\theta}^{3600} \approx  1.3634\pm 0.0005.
  \end{dmath}
  
  
  \textit{The disjoint transition.}. Our computational results show that magnetic transition \eqref{eq:critical_J_magnet_2D} and structural transition \eqref{eq:critical_J_theta_2D} appears at the different points. The intervals within errobars do not overlap. Therefore, we cannot conclude that magnetic phase transition and structural transition happens at the same point, in contrast to Ising model on SAWs. Our MC the data is inconclusive, whether the transitions occur simultaneously or at distinct values of the coupling constant J. More work is needed to conclusively rule out one of possibilities.

  \subsection{Distribution of $\langle\cos \theta \rangle$ and $\langle e \rangle$ } \label{sec:distributions}
  To study the phase transition order, we look at distributions of energy and magnetization.
  
  \begin{figure}[ht!]
  	\centering 
  	\includegraphics[scale=0.25]{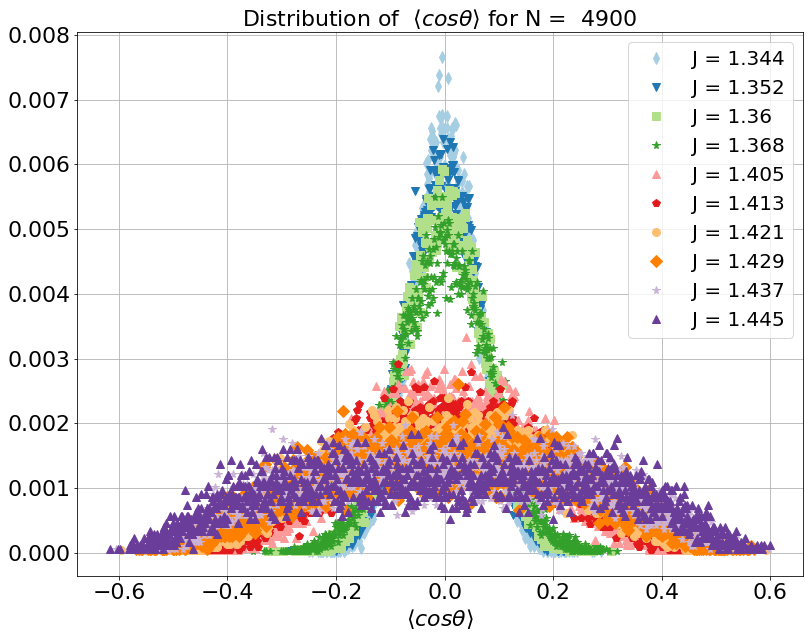}
  	\includegraphics[scale=0.25]{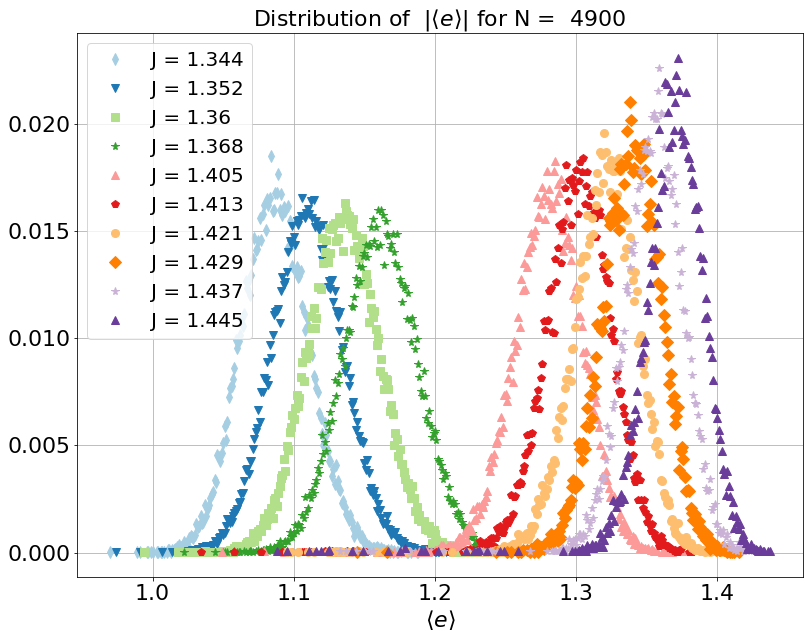}
  	\caption{ Distributions of mean energy and mean $\cos \theta$ as magnetization component for chains $N=4900$ for various $J$ over structural and magnetic transitions.    }
  	\label{fig:distributions}
  \end{figure}

  Figure \ref{fig:distributions} shows shapes of energy distribution across the structural transition \eqref{eq:critical_J_theta_2D} and magnetic transition  \eqref{eq:critical_J_magnet_2D}. The curves of distributions is Gaussian-like and do not sign any bimodal shapes. 
  
  Additionally to energy, we consider mean $\cos \theta$ which represents a component of mean magnetization vector \eqref{meanmagnetization}. For points $J < J_{cr}$, curves of $\cos \theta$ distribution is similar to normal curve. Similarly to study of Ising model on SAWs \cite{PhysRevE.104.024122}, we check signs of phase coexistence or not. The shape of magnetization distribution reflects phase coexistence as trimodal shape, where left and right modals corresponds to the "ordered states" and the central one comes from "disordered" states. 
  Here we  see no signs of phase coexistence, which is consistent with a continuous transition.

  \section{Numerical simulations, 3D case} 
  
  In this section, we consider short chains up to $N=700$ which is much shorter than we study for 2D case. The reason of it is the lattice implementation which requires to keep $2dN^3=6N^3$ nodes in the memory.

    \subsection{Structural properties}

    \begin{figure} 
    	\centering
    	\includegraphics[scale=0.22]{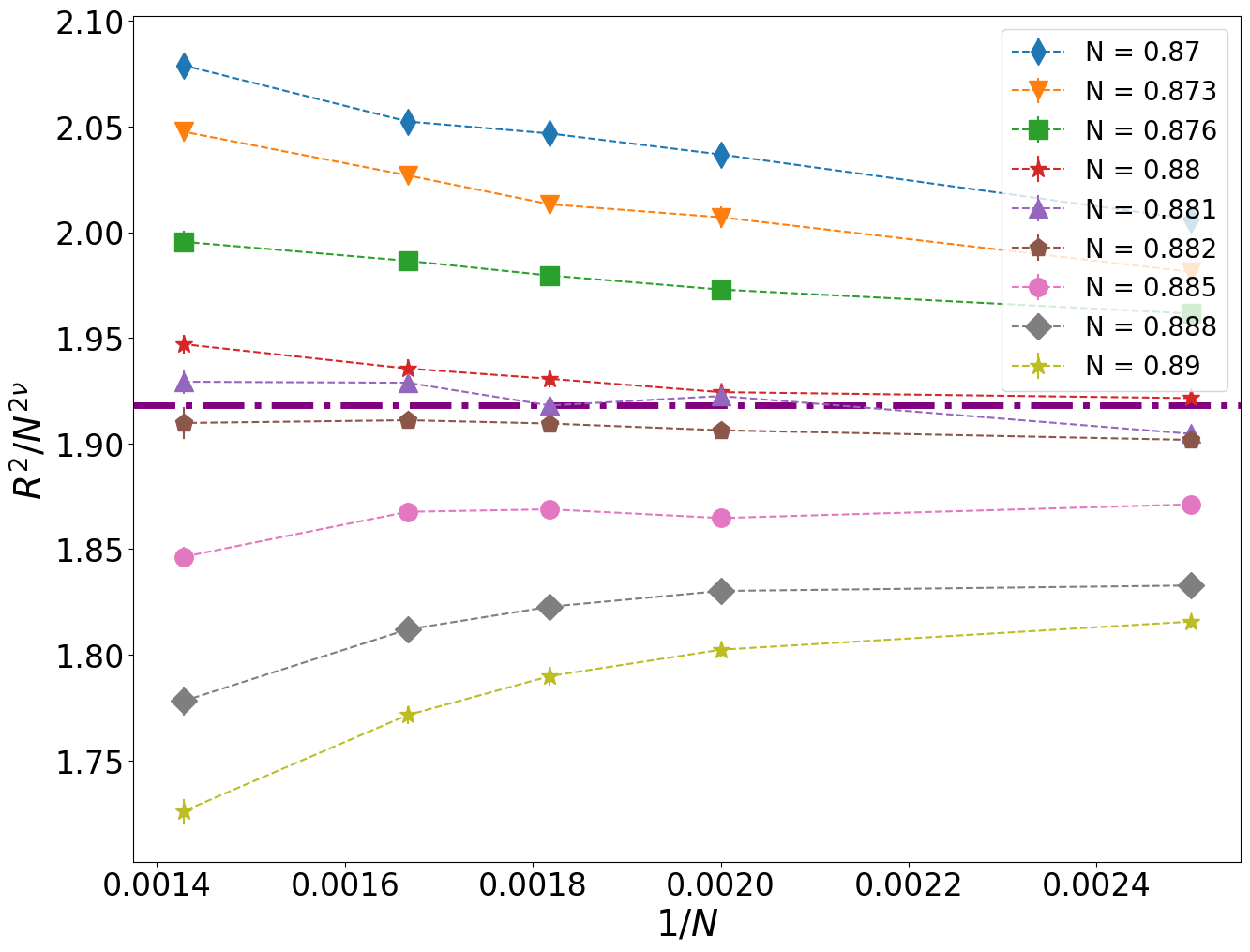} 
    	\caption{  Scaled Mean-squared end-to-end distance using $\nu_{\theta}=1/2$ as a function of $1/N$ from $N=400$ to $N=700$.  The dash-dotted  purple horizontal line corresponds to structural phase transition and placed at the estimation \eqref{eq:critical_J_theta_3D} obtained paired regression method from Sec. \ref{pairedregressions}.  }
    	\label{fig:Rscaled3D}
    \end{figure}
    
    First, we focus on  structural properties and structural transition. Following 2D case, we expect that the XY model on SAWs inherits the critical value parameter $\nu=1/2$ from interacting SAWs. Figure \ref{fig:Rscaled3D} shows the scaled
    mean-squared end-to-end distance by $\nu_{\theta}=1/2$ \eqref{eq:nu_3D_Jtheta_flory} as a function of the chain length
    $1/N$ for a range of $J$. The horizontal line is expected to represent the point of structural phase transition and corresponds to the critical exponent $\nu_{\theta}$ \eqref{eq:nu_3D_Jtheta_flory}.  We place the horizontal line at the estimated value where scaled curves cross using histogram procedure described in Sec. \ref{pairedregressions}. The horizontal line is limited by red star-marked curve ($J=0.88$)  and pink circle-marked curve ($J=0.885$).  Therefore, according our calculations for chains up to $N=700$, the system undergoes the structural phase transition $J \in [0.88;0.885] $ with the critical exponent $\nu_{\theta} = \frac{1}{2}$. However, this visual inspection is not very reliable due to finite size effects.  
  Applying paired regressions, we obtain the following estimate for critical interaction energy: 
    \begin{dmath}
    \label{eq:critical_J_theta_3D}
    J_{\theta}^{700} \approx 0.876(5) .   
    \end{dmath} 
    
    For further study, we note the critical exponent value $\nu_{\theta} = \frac{1}{2}$ and use it in the following section to scale mean radius.

      \subsubsection{Magnetic phase transition}
    
    To investigate the critical behaviour at the phase transition, we again calculate  the Binder cumulants values \eqref{binderqum} and scaled mean end-to-end distance  \eqref{endtoend}.
    \begin{figure}
    	\centering
    	\captionsetup{justification=centering}
   
    	\begin{subfigure}[b]{0.45\textwidth}
    		\includegraphics[width=\textwidth]{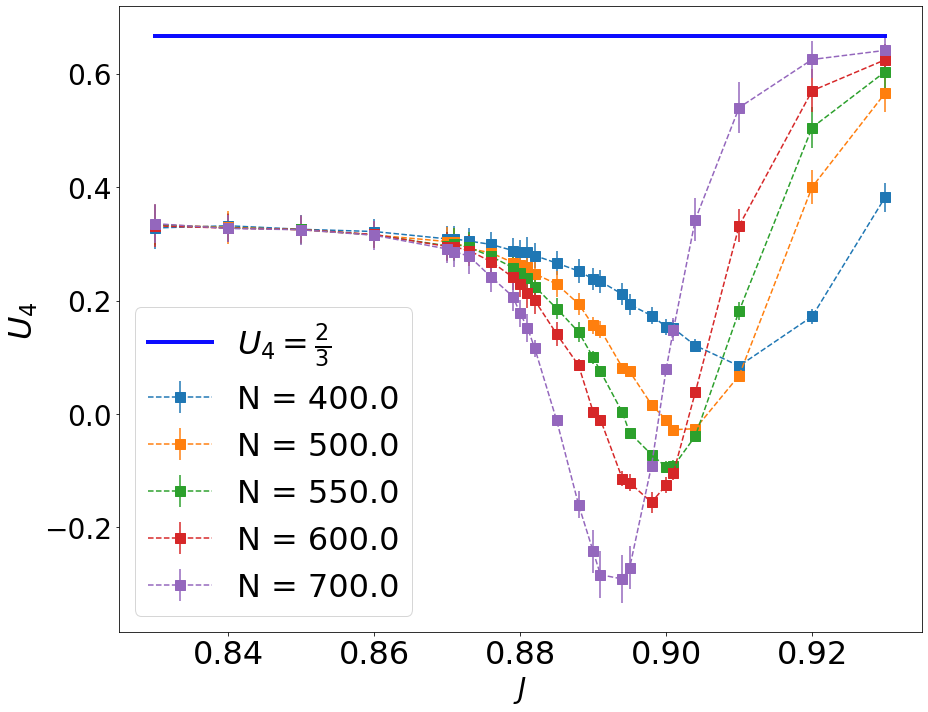}
    		\caption{Binder cumulant on the interval including magnet phase transition.}
    		\label{fig:bcshort_longbc_3}
    	\end{subfigure}
    	
    	\begin{subfigure}[b]{0.45\textwidth}
    		\includegraphics[width=\textwidth]{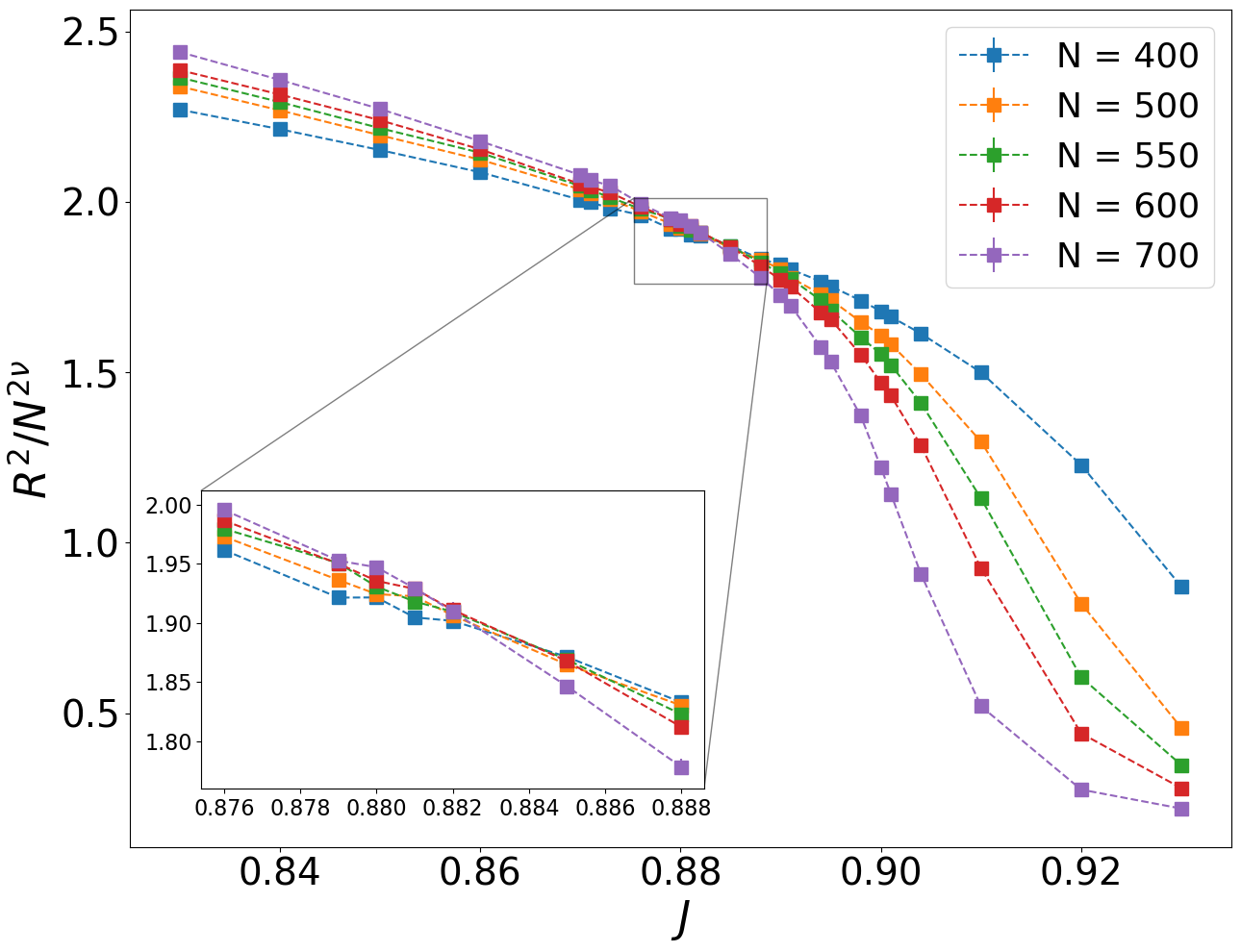}
    		\caption{Scaled mean radius for long chains on the region with structural phase transition. }
    		\label{fig:bcshort_longradius_3}
    	\end{subfigure}
    	\caption{$h=0$. Binder  cumulants \eqref{binderqum} and mean radius \eqref{endtoend}. The mean radius is scaled by $\nu = \frac{1}{2}$.   }
    	\label{fig:bcshort_3D}
    \end{figure}

   The Binder parameters (top Figure \ref{fig:bcshort_3D}) has lower limiting value $U_4=\frac{1}{3}$ as $J\rightarrow0$ (see Sec. \ref{u4xy} for more details). For large values of coupling constant $J$, the system gets the ordering phase, for which $U_4=\frac{2}{3}$. 
    
    Cumulant curves  diverges at the critical region. This is a clear sign of a first-order phase transition \cite{PhysRevLett.47.693, PhysRevB.30.1477}. The similar results for diverged cumulants were obtained for Ising model on SAWs for 3D case \cite{PhysRevE.104.024122}.

     \subsection{Distribution of $\langle\cos \theta \rangle$ and $\langle e \rangle$ } \label{sec:distributions_3D}

        We check the distributions of thermodynamic characteristics to check whether energy distribution is bimodal and magnetic distribution show signs of phase coexistence.
    
    \begin{figure}
    	\centering
    	\captionsetup{justification=centering}
    	\begin{subfigure}[b]{0.45\textwidth}
    		\includegraphics[width=\textwidth]{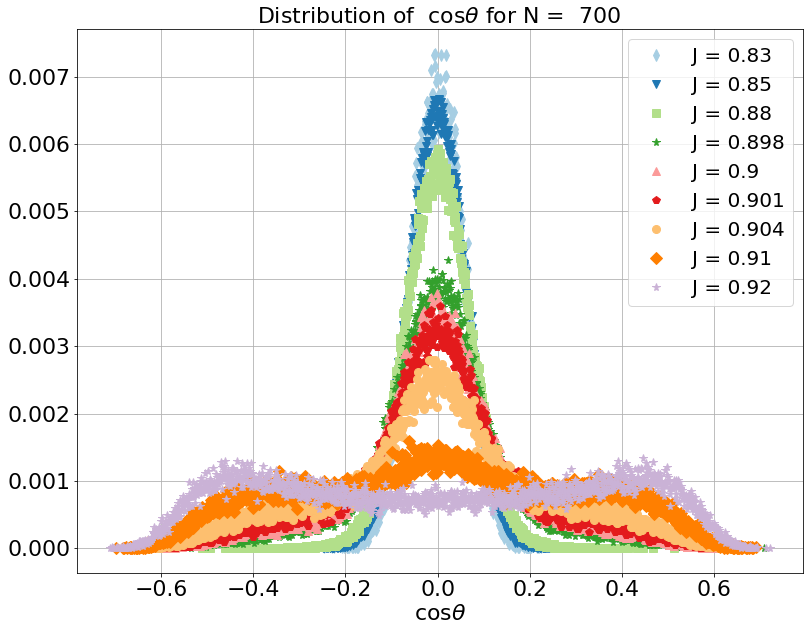}
    		\centering{\caption{ The mean component $\cos \theta$ of magnetization vector.   } }
    	\end{subfigure}
    	\begin{subfigure}[b]{0.45\textwidth}
    		\includegraphics[width=\textwidth]{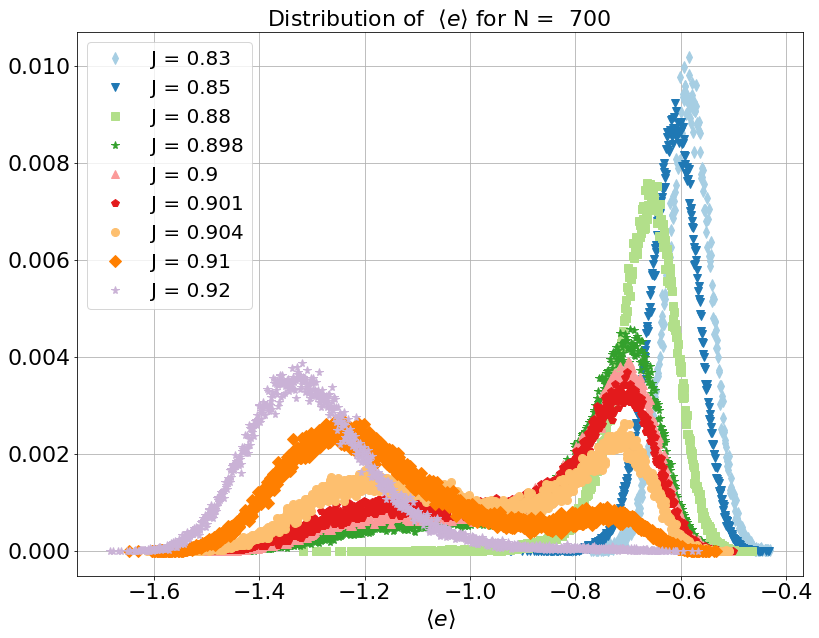}
    		\caption{ The mean energy.  }
    	\end{subfigure}
    	\caption{ Distributions for chain $N=700$ for various $J$.  }
    	\label{fig:Distributions_3D}
    \end{figure}
    
    Figure \ref{fig:Distributions_3D} (top) illustrates that energy distribution has bimodal shape approximately at $J \approx 0.904$. However, this region of bimodal curve is quite far ($\approx 0.03$) from the estimation for point of the structural transition \eqref{eq:critical_J_theta_3D} and the divergence region of minimum Binder cumulant (see Figure \ref{fig:bcshort_longbc_3}). This could be caused by finite size effect as chains up to $N=700$ are not too long. 
    
    We consider mean $\cos \theta$ which is a component of mean magnetization vector \eqref{meanmagnetization}. For points $J < J_{cr}$ before magnetic transition
    transition, curves of $\cos \theta$ distribution is similar to normal curve which
    is expected as this case corresponds to the sampling from uniform distribution $U \sim [-1;1]$ and convergence to the Gaussian.Over the critical region, the shapes of distributions are far from Normal-like curves.

 


\section{Conclusions and outlook}

We study the XY model on self-avoiding walks on a square lattice in 2D and a simple cubic lattice in 3D
using Monte-Carlo simulations. We consider the regime where both spins and SAW conformations are fluctuating---in a sense, this is an XY model defined on a dynamic lattice with annealed disorder.

We use unbiased Monte-Carlo simulations which combine a variant of the canonic-ensemble worm algorithm for conformations and cluster updates for spin variables. This way, our simulations keep being efficient in the critical region around the $theta$-point, and allow us to simulate SAW chains of up to $5\times 10^3$ in 2D and up to $7\times 10^2$ in 3D.

In 2D, our numerical results indicate that both paramegnetic-to-ferromagnetic and globule-coil transitions are continuous. This is consistent with earlier results for a similar model with Ising spins \cite{PhysRevE.104.054501,PhysRevE.104.024122}. The structural and magnetic transitions occur in the same region of the coupling constant $J$, but the numerical values of $J$ differ within statistical errorbars. This is in contrast to the Ising case, where the transitions occur simultaneously  \cite{PhysRevE.104.054501,PhysRevE.104.024122}.
A possible explanation is that we are limited by the finite size effects even for the SAW chains of $\sim 10^3$ sites---for the Kosterlitz-Thouless type transition of the 2D XY model, the correlation length has an exponential scaling, and up to $\sim 10^6$ spins were necessary to accurately resolve the KT physics in previous Monte-Carlo simulations \cite{Hasenbusch_2008}. 

Our numerical simulations indicate that the mean end-to-end distance (equivalently, the gyration radius) of a SAW in the critical region scales with the length of the SAW as $\langle R^2 \rangle \sim N^{2\nu}$, with the value of $nu = 4/7$, inherited from the classic interacting SAW model \cite{Duplantier1987}. We note that this same scaling is observed for the Ising model \cite{PhysRevE.104.054501,PhysRevE.104.024122} and for the dynamic HP model \cite{Faizullina2021}. We thus speculate that this critical exponent is inherited by a wide class of models irrespective of details of short-range interactions between monomers of a SAW.

In 3D, our simulations are limited to the SAW lengths of up to $~700$, which is an almost order of magnitude smaller than 2D. The limitation is purely technical and is due to our implementation \cite{gitsaw} using locally sensitive hashing to achieve $O(1)$ time complexity for the nearest-neighbor queries. Despite this limitation, the SAW lengths available in our simulations are sufficient for drawing quantitative conclusions from the MC data.
Specifically, we see clear signatures of the magnetic transition being first order: the Binder cumulant displays a characteristic divergence (of course, the divergence is strongly rounded by the finite-size effects) and the distributions of observables signal phase coexistence. The critical exponent for the gyration radius is consistent with the 3D interacting SAW value $\nu = 1/2$ \cite{van2015statistical}. We note that this scenario is similar to the one observed for the Ising model on a 3D SAW in Refs  \cite{PhysRevE.104.054501,PhysRevE.104.024122}: the magnetic transition is first order and the gyration radius exponent is consistent with the interacting SAW value. Therefore, we speculate that this scenario is generic and should be observed for a wide range of SAW models with short-range interactions. Whether long-range interactions (e.g. dipole-type $\sim 1/r^3$ couplings) change the behavior is an open question.

Numerical simulations were performed using the  computational
resources of HPC facilities at HSE University \cite{Kostenetskiy_2021}.

\section{Appendix } \label{appendix}
\subsection{$U_4$ as $ J\rightarrow 0$} \label{u4xy}

Consider the case when no interaction which could be close to classical 1-dimensional XY-chain. In case of open boundary conditions, the partition function for the chain of the length $N$ has following form:  
\begin{dmath}
\label{partitionfunction_free}
Z(J) =    \int_{-\pi}^{\pi} \frac{1}{ (2 \pi  )^N}       d \theta_1 d \theta_2 \dots d\theta_N
e ^{J\cos(\theta_1-\theta_2)} e ^{J\cos(\theta_2-\theta_3)} \\ \dots 
e ^{J\cos(\theta_{N-1}-\theta_N)} 
\end{dmath}

In case $J=0$ (high-temperature regime), all states have equal probabilities:
\begin{dmath}
\label{paritionfunction_free_zero}
Z(0) =  
\int_{-\pi}^{\pi} (\frac{1}{2 \pi})^N   d \theta_1 d \theta_2 \dots d\theta_N 
\end{dmath}
To calculate the exact value of $\langle m^2 \rangle (J=0)$  we use following results: 
\begin{dmath*} 
\int_{-\pi}^{\pi}  \frac{1}{2 \pi} \sin^2 \theta d \theta =\int_{-\pi}^{\pi}  \frac{1}{2 \pi} \cos^2 \theta d \theta = \frac{1}{2} 
\end{dmath*}
\begin{dmath*} \int_{-\pi}^{\pi}  \frac{1}{2 \pi}\sin \theta d \theta  =\int_{-\pi}^{\pi}  \frac{1}{2 \pi}\cos  \theta d \theta = 0 \end{dmath*}
After some calculation, only integration results for $N$ times $sin^2 \theta_i$ and $N$ times $cos^2 \theta_i$ survive:
\begin{dmath} 
\label{m2j0}
\langle m^2 \rangle (J=0) = \frac{1}{N^2}   \int_{-\pi}^{\pi}  (\frac{1}{2 \pi})^N \left(  
( \sum_{i=1}^{N}\cos \theta_i )^2 +  (\sum_{i=1}^{N}\sin \theta_i  )^2 \right)  d \theta_1 d \theta_2 \dots d\theta_N = \\
\frac{1}{N^2}  (\frac{1}{2}N +\frac{1}{2}N)   = \frac{1}{N}
\end{dmath}
Next, to calculate $\langle m^4 \rangle (J=0)$ we use following facts: 
\begin{dmath*}
\langle m^4 \rangle (J=0) = \frac{1}{N^4}   \int_{-\pi}^{\pi}  (\frac{1}{2 \pi})^N \left(  
( \sum_{i=1}^{N}\cos \theta_i )^2 +  (\sum_{i=1}^{N}\sin \theta_i  )^2 \right)^2 d \theta_1 d \theta_2 \dots d\theta_N  
\end{dmath*} 
\begin{dmath*}
\left(  
( \sum_{i=1}^{N}\cos \theta_i )^2 +  (\sum_{i=1}^{N}\sin \theta_i  )^2 \right)^2 =  (\sum_{i=1}^{N}\cos \theta_i )^4 + (\sum_{i=1}^{N}\sin \theta_i )^4 + 2 ( \sum_{i=1}^{N}\cos \theta_i )^2( \sum_{i=1}^{N}\sin \theta_i )^2 
\end{dmath*} 

$ \int_{-\pi}^{\pi}  \frac{1}{2 \pi}\sin^4 \theta d \theta =\int_{-\pi}^{\pi}  \frac{1}{2 \pi}\cos^4 \theta d \theta = \frac{3}{8}$ (We  have $N$ times $sin^4\theta_i$-term and $N$ times $cos^4\theta_i$-term  what results in $2\times \frac{3}{8} \times N$). \\

$ \int_{-\pi}^{\pi}  \frac{1}{2 \pi}\int_{-\pi}^{\pi} \frac{1}{2 \pi}\sin^2 \theta_i\sin^2 \theta_j d\theta_i  d\theta_j = \int_{-\pi}^{\pi}  \frac{1}{2 \pi}\int_{-\pi}^{\pi} \frac{1}{2 \pi}\cos^2 \theta_i\cos^2 \theta_j d\theta_i  d\theta_j = \frac{1}{4}$ (We  have $6N(N-1)\frac{1}{2}$ times $sin$-term and $6N(N-1)\frac{1}{2}$ times $cos$-term  what results in $6\times \frac{1}{4} \times N(N-1)$). \\

$ \int_{-\pi}^{\pi}  \frac{1}{2 \pi}\sin^2 \theta\cos^2 \theta d \theta  = \frac{1}{8}$ (We  have this term $2N$ times  what results in $2\times \frac{1}{8} \times N$). \\

$ \int_{-\pi}^{\pi}  \frac{1}{2 \pi}\int_{-\pi}^{\pi} \frac{1}{2 \pi}\cos^2 \theta_i\sin^2 \theta_j d\theta_i  d\theta_j = \frac{1}{4}$ (We  have $2N(N-1)\frac{1}{2}$ times  what results in $2\times \frac{1}{4} \times N(N-1)$). \\

All other terms with odd power of sin and cos function equals zero after integration over period. 
\begin{dmath*}
\langle m^4 \rangle (J=0) = \frac{1}{N^4} \left( 2\times \frac{3}{8} \times N + 6\times \frac{1}{4} \times N(N-1) + 2\times \frac{1}{8} \times N + 2\times \frac{1}{4} \times N(N-1)
\right) = \frac{2N-1}{N^3}  
\end{dmath*}
\begin{dmath}
\label{binderqum_0}
U_4 (J=0) = 1 - \frac{  \frac{2N-1}{N^3} }{3  \frac{1}{N^2} } \\ = 1 - 
\frac{2N-1}{3N} = \frac{1}{3} + \frac{1}{3N} 
\end{dmath}


\subsection{Paired regressions} \label{pairedregressions}

 To estimate critical values for cumulants $U_4$ and phase transition point $J_{cr}$, we perform paired linear intersections. The procedure to analyze Monte-Carlo data is following: 
 
 1. Choose the pair of two different N values for length of the chain. Choose the range of values for interaction energy J. This segment should be as short as possible and include the point of intersection of the two curves. 
 
 2. We need to obtain the errors to estimated Binder cumulant. To that end, we use Gaussian sampling. 
 
 For each point from the set generate $n_{samples}$ values using Normal distribution with mean and standard error of $\langle m^2 \rangle  $ and $\langle m^4 \rangle$ as parameters: $ M2_{J, N} \sim N (\langle m^2 \rangle  , \sigma (\langle m^2 \rangle  ))$, $ M4_{J, N} \sim N (\langle m^4 \rangle  , \sigma (\langle m^4 \rangle  ))$. We generate for each value $J$ 1000 samples. For each pair of sampled $m_2, m_4$ we calculate the Binder cumulant \eqref{binderqum}. 
 
 3. Using generated set, for each pair $J$ and $N$ make estimation for mean and standard deviation $\langle U_4 \rangle$.
 
 4. Now, we have  two curves of calculated $U_4$ with errorbars for two values of $N$. Apply weighted least squares regression to find crossing point. Save the obtained estimation for $\hat{J}$. 
 
 5. Repeat steps 2-5 $n_{lines}$ times. We repeat it $n_{lines}=1000$ times. 
 
 6. At the end, we have $n_{lines}$ of estimated $\hat{J}$ and $U_{4\mathrm{critical}}$ where two curves cross. The mean value and standard deviation of this arrays correspond to the estimation and its error.

The same procedure could be applied using crossing curves of $R^2/N^{2\nu}$ to estimate $J_{\theta}$ and crossover value for $R^2/N^{2\nu}$.

\bibliography{bibliography}

\end{document}